# Lateral heterostructures formed by thermally converting n-type SnSe$_2$ to p-type SnSe

*Zhen Tian,[1,2,4,†] Mingxing Zhao,[2,4,†] Xiongxiong Xue,[3,†] Wei Xia,[2] Chenglei Guo,[1,2,4] Yanfeng Guo,[2] Yexin Feng[3]\* and Jiamin Xue[2,4,5,]\**

[1]Shanghai Institute of Optics and Fine Mechanics, Chinese Academy of Sciences, Shanghai 201800, China

[2]School of Physical Science and Technology, ShanghaiTech University, Shanghai 201210, China

[3]School of Physics and Electronics, Hunan University, Changsha 410082, People's Republic of China

[4]University of Chinese Academy of Sciences, Beijing 100049, China

[5]Center for Excellence in Superconducting Electronics (CENSE), Chinese Academy of Sciences, Shanghai 200050, China

\*xuejm@shanghaitech.edu.cn

\*yexinfeng@pku.edu.cn

ABSTRACT

Different two-dimensional materials, when combined together to form heterostructures, can exhibit exciting properties that do not exist in individual components. Therefore, intensive research efforts have been devoted to their fabrication and characterization. Previously, vertical and in-plane two-dimensional heterostructures have been formed by mechanical stacking and chemical vapor deposition. Here we report a new material system that can form in-plane p-n junctions by thermal conversion of n-type $SnSe_2$ to p-type SnSe. Through scanning tunneling microscopy and density functional theory studies, we find that these two distinctively different lattices can form atomically sharp interfaces and have a type II to nearly type III band alignment. We also demonstrate that this method can be used to create micron sized in-plane p-n junctions at predefined locations. These findings pave the way for further exploration of the intriguing properties of the $SnSe_2$-SnSe heterostructure.

INTRODUCTION

P-n junctions lie at the heart of solid state electronics. Modern silicon based devices rely on ion implantation to convert lithographically defined patterns of either p or n type wafers to the opposite doping type, forming p-n junctions.[1] However, ion implantation has limited application in newly emerged two-dimensional (2D) electronic materials. Due to their many novel properties, 2D materials have attracted a lot of research efforts in the last few years.[2-4] The family of 2D materials has been ever enlarging, ranging from superconductors (*e.g.* NbSe$_2$),[5,6] semimetals (*e.g.* graphene),[7] semiconductors (*e.g.* MoS$_2$),[8] to insulators (*e.g.* BN). Among them, the semiconducting 2D materials, such as the transitional-metal dichalcogenides (TMDs), received special attention for their possible applications in electronics. With only a few atomic layers, this group of materials shows high mobility,[9] suitable band gaps,[10] intriguing valley physics[11] and so on. As building blocks for complicated devices, p-n junctions based on 2D materials are highly demanded. Ion implantation, the traditional method for forming p-n junctions in three-dimensional silicon, cannot be used here due to the atomically thin body of 2D materials. So far, mainly two routes have been taken to tackle this problem. In one way, mechanical exfoliation and sequential stacking of naturally p and n type 2D flakes were used to form vertical heterojunctions. Light emitting diodes,[12] tunnel devices[13] and photovoltaics[14] have been realized with devices fabricated this way. Alternatively, chemical vapor deposition (CVD) was used to obtain in-plane heterostructures, which are preferred for their planer geometry. With this method, in-plane heterostructures and p-n junctions consisting of WS$_2$/WSe$_2$,[15] WS$_2$/MoS$_2$,[16] MoSe$_2$/WSe$_2$,[17] WSe$_2$/MoS$_2$[18] and MoS$_2$/MoSe$_2$[19] were successfully grown and characterized. A most recent work has further improved this method to a new level and in-plane TMD superlattice can be grown.[20] However, all these heterojunctions are based on CVD grown TMDs. It is interesting and important to explore other 2D systems that can form planer p-n junctions with new mechanisms. In this article, we report a novel method for creating in-plane p-n heterostructures by selective thermal conversion of layered SnSe$_2$ to SnSe. We use Raman spectroscopy to verify the coexistence of SnSe$_2$ and SnSe after thermal annealing of single crystalline SnSe$_2$ in vacuum. Then we use scanning tunneling microscope (STM) to study the heterostructure at the atomic level. Unlike previously reported TMD heterostructures, which are all formed by hexagonal lattices, the SnSe$_2$-SnSe system consists of a hexagonal and an orthorhombic lattice. Surprisingly, our STM results show that the two distinct lattices can form atomically sharp interfaces. Further scanning tunneling spectroscopy (STS) measurements show that SnSe$_2$ is of intrinsic heavy n-type doping, while SnSe is of intrinsic heavy p-type doping. Their junction forms a type II to nearly type III broken-band alignment. Density functional theory (DFT) calculations find that defects such as Se (Sn) vacancies and interstitial Sn (Se) would result in the n (p) type doping in SnSe$_2$ (SnSe). Among them, the interstitial Sn (Sn$_i$) in SnSe$_2$ and Sn vacancies (V$_{Sn}$) in SnSe are the dominant dopants due to the lower formation energies of them. Simulated 2D contours of the electron partial charge density match very well with the experimental STM images. Density of states (DOS) curves from the DFT calculations also confirm the type II to nearly type III broken-band alignment of the SnSe$_2$-SnSe junction. In addition to the atomic scale study by STM and DFT, we demonstrate that micrometer sized SnSe$_2$-SnSe junctions can be controllably produced in few-layer SnSe$_2$ flakes. This method of forming in-plane p-n junctions by thermal conversion and its comprehensive microscopic study is of great importance for creating and studying new heterostructures based on this system.

SnSe$_2$ belongs to the hexagonal crystal system, with the space group of P$\bar{3}$m1. It crystallizes in the 1T layered structure, as shown in Fig. 1a. Unlike the common 2H structure of TMDs, each layer of SnSe$_2$ has the inversion symmetry and six selenium atoms occupy the octahedral coordination around a tin atom. Optical absorption data has indicated that bulk SnSe$_2$ has an indirect band gap of about 1 eV.[21,22] Being a typical van der Waals material, it can be readily exfoliated or CVD grown into few layer flakes. Its electronic and photoelectronic properties as a 2D material have been studied by several groups. Transport studies found it to be a heavily n doped material with electron density up to $1.3 \times 10^{19}$ cm$^{-3}$.[23] Few-layer SnSe$_2$ field-effect transistors (FETs) without any device optimization already showed a room temperature mobility to be as high as 85 cm$^2$V$^{-1}$s$^{-1}$,[23] compared favorably with that of the TMDs. Photoelectronic devices made on monolayer SnSe$_2$ also showed its superior photosensitivity.[24] On the other hand, SnSe belongs to the orthorhombic lattice system, with the space group Pnma.[25,26] As a newly discovered thermal electric material with the record high figure of merit ZT,[27,28] bulk SnSe has been studied extensively in the last few years.[29-33] 2D SnSe flakes have been synthesized with CVD and their electrical properties were explored.[34,35] Transport measurements showed that SnSe is intrinsically p doped with interesting in-plane anisotropy.[35] Besides the experimental efforts, 2D SnSe also attracted a lot of theoretical interests[36-40] due to their puckered lattice structure (Fig. 1a), which is similar to that of black phosphorus. Properties such as tunable in-plane ferroelectricity,[41] giant piezo-electricity,[37] valley physics[40] and so on have been predicted.

RESULTS AND DISCUSSION

Previous transmission electron microscopy (TEM) study has shown that thin flakes of SnSe$_2$ can be partially converted to nanometer sized patches of SnSe with high energy electron beam radiation.[42] This result showed that Se atoms in SnSe$_2$ can be removed with external energy input. However, to be compatible with common device fabrication process, thermal conversion is preferred. And also, a technique that can probe both the crystal structure and local electronic properties of the SnSe$_2$-SnSe system is needed to study the band alignment at the junctions. To this end, we use the low temperature ultrahigh vacuum STM (Omicron LT) equipped with *in situ* annealing capability. Fig. 1b shows the SnSe$_2$ surface imaged by STM before annealing. The triangular lattice of Se atoms is clearly visible and defects are seldom observed. The fast Fourier transform (FFT) of the topography image is shown in Fig. 1c. The measured lattice constant is 0.385 nm, consistent with literature values obtained from X-ray diffraction[43] and our calculated lattice constants (see the Supporting Information, SI).

To explore the possibility of thermally converting SnSe$_2$ to SnSe, we anneal the STM scanned SnSe$_2$ crystal in the STM chamber. We found that through annealing the SnSe$_2$ can be completely transformed to SnSe. By carefully tuning the temperature and annealing time (see Experimental Section, ES, for details), we can achieve a partial conversion, which is confirmed with Raman spectroscopy as shown in Fig. 1d. We can see that Raman peaks corresponding to both SnSe$_2$ ($A_{1g}$ mode at 182 cm$^{-1}$) [44] and SnSe ($A_g$ mode at 65 cm$^{-1}$, 125 cm$^{-1}$, 148 cm$^{-1}$; $B_{3g}$ mode at 105 cm$^{-1}$)[45] vibrational modes are present in the data. Since the laser spot size used to obtain the Raman spectrum is about 1 μm, this indicates that the SnSe$_2$ single crystal has turned into a polycrystalline sample with submicron sized SnSe$_2$ and SnSe crystalline patches.

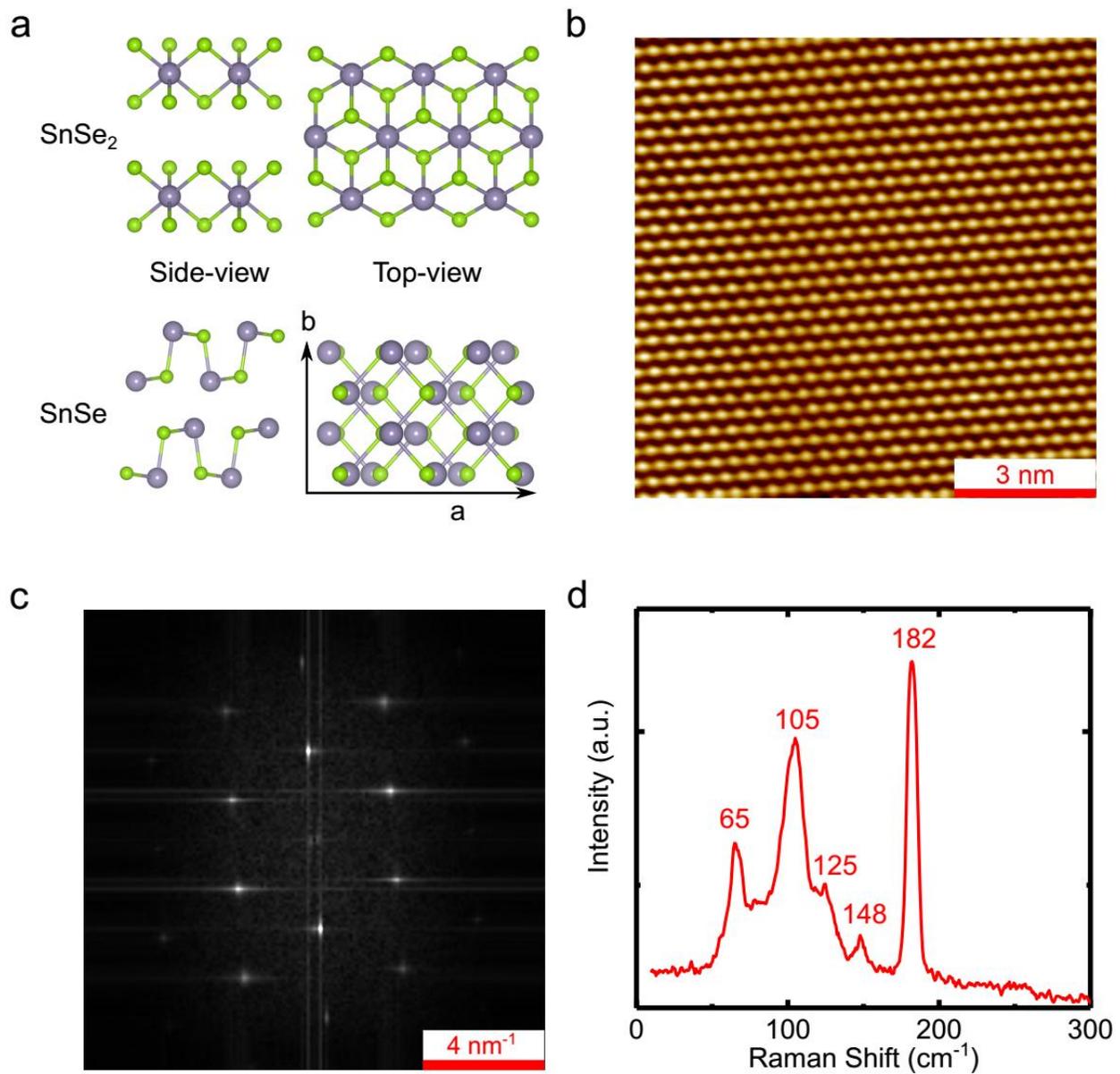

**Figure 1.** Basic characterization of SnSe$_2$. (a) Atomic model of bulk SnSe$_2$ (upper) and SnSe (lower). Green and violet balls are the Se and Sn atoms, respectively. (b) Constant current STM topography image of single crystalline SnSe$_2$ with sample bias voltage $V$ = 1.5 V and current setpoint $I$ = 100 pA. (c) FFT of (b). (d) Raman spectrum of annealed SnSe$_2$.

The Raman result is confirmed by the STM images. As shown in Fig. 2a, after annealing we can find atomically flat areas (usually with a size of the tens to hundreds of nanometers) separated by some high ridges (the white part in Fig. 2a). The lower part of Fig. 2a is clearly a triangular lattice, while the upper part shows a hint of an orthorhombic lattice. When zooming in to each individual flat area with optimized imaging parameters, we can nicely resolve both lattices, as shown in Fig. 2b and 2c, corresponding to the upper and lower part of Fig. 2a, respectively. From measuring the lattice constants in both the real space image and the FFT pattern, we identify that the upper (lower) part is indeed SnSe ($SnSe_2$). The measured lattice constants of the orthorhombic SnSe are a = 0.439 nm and b = 0.419 nm,[46,47] and that of the hexagonal $SnSe_2$ is 0.383 nm, both close to the calculated lattice constants with the DFT methods (in the SI). According to the atomic model shown in Fig. 1a, the top layer in SnSe consists of both tin and selenium atoms, with the tin atoms slightly higher. Tin vacancies are commonly observed as shown in Fig. 2b, while the $SnSe_2$ surface retains its perfect lattice. Besides this type of rough-ridge interfaces between SnSe and $SnSe_2$, we also find many one-atomic-layer-step interfaces, in which the top layer may be SnSe or $SnSe_2$ and the bottom layer be the other (Fig. S1). Most interestingly, we find that $SnSe_2$ and SnSe can form atomically sharp interfaces, as shown in Fig. 2d. The left half is $SnSe_2$ with three-fold symmetry, while the right half is the orthorhombic SnSe. Previously reported in-plane heterostructures can also have atomically sharp interfaces. But they were all formed between hexagonal lattices with similar lattice constants.[15-20] Here we see that the two materials with distinct lattice constants and crystal symmetries can be connected seamlessly with the transition width of only one bond length. The possible bond arrangement connecting the two materials will be discussed in the theoretical section. It is also interesting to note that the lattice of $SnSe_2$ has the same orientation before (Fig. 1b) and after (lower part of Fig.

2a, Fig. 2c and left part of Fig. 2d) annealing, while that of the newly formed SnSe is rotated related to $SnSe_2$. Through extensive scanning (Fig. S2 in SI), we find that SnSe can only have specific rotation angles related to $SnSe_2$. If we define a Cartesian coordinate as shown in Fig. 2e, the possible orientations, differing by 15 degrees (with the experimental error of $\pm 1.3$ degree), are shown by the dotted lines in Fig. 2e. This is probably due to the formation kinetics of SnSe from $SnSe_2$. Previous study of electron beam induced conversion of SnSe2 to SnSe (see Ref. 42) proposed a possible formation path. At the first stage electron beam would induce atomic defects of Se and then as the density increases, these defects tend to cluster together due to the attracting potential between them. When the density reaches a critical value, the original $SnSe_2$ crystal structure becomes unstable and the new crystal of SnSe forms. From this proposed formation path, the $SnSe_2$ crystal orientation would have a strong confinement on the formed SnSe, since the crystalline SnSe "emerges" from the $SnSe_2$ lattice. In our method, we believe that thermal energy played the role of high energy electron beam in inducing defects and the formation path is similar.

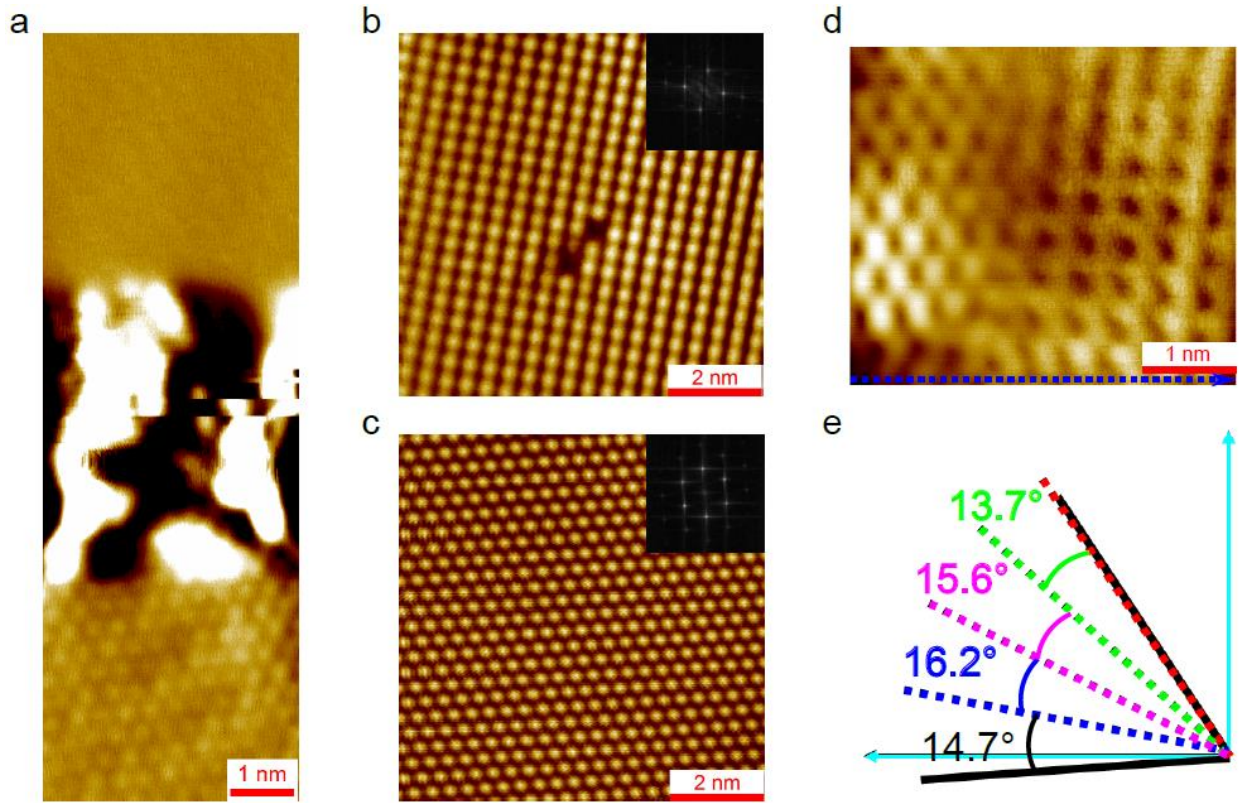

**Figure 2.** STM topography of different regions on annealed $SnSe_2$. (a) Large flat area of $SnSe_2$ and SnSe are both present with a very rough interface. (b) SnSe area of the upper region in (a). Inset: the corresponding FFT pattern. (c) $SnSe_2$ area of the lower region in (a). Inset: the corresponding FFT pattern. (d) The atomically sharp interface between $SnSe_2$ and SnSe. STM scanning parameters: (a, c, d) $V = 1.5$ V, $I = 50$ pA; (b) $V = 1.5$ V, $I = 100$ pA. (e) The relationship between $SnSe_2$ crystal orientation (black solid lines) and the thermally-converted SnSe crystal orientations (dotted lines, corresponding to the zigzag direction). The angle between each two lines is about 15°.

The STM topography can reveal the atomic structures. However, to obtain the electronic structures and band alignment between $SnSe_2$ and SnSe, the STS technique is needed. At zero temperature, the tunneling current $I$ can be expressed as[48]

$$I = \frac{2e}{h} \int_0^{eV} \rho_T(E_F - eV + \epsilon)\rho_S(E_F + \epsilon)|M(\epsilon)|^2 d\epsilon,$$

where $e$ is the electron charge, $h$ the Plank constant, $V$ the bias voltage applied to the sample related to the tip, $E_F$ the Fermi energy of the tip-sample system at thermal equilibrium, $\rho_T$ ($\rho_S$) the DOS of the tip (sample) and $M$ the tunneling matrix. For an approximately energy independent $\rho_T$ (confirmed in our experiment by checking the STS spectrum against the surface state on Ag (111) before and after scanning on the SnSe$_2$-SnSe sample), we get

$$\frac{dI}{dV} \propto \rho_S(E_F + eV)|M(eV)|^2,$$

so the d$I$/d$V$ is proportional to the sample DOS and the tunneling matrix. We first obtain STS spectra on large flat areas (lateral dimension in the hundreds of nanometers) of SnSe$_2$ and SnSe as shown in Fig. 3a and 3b, respectively. Similar spectra have been repeatedly obtained over 40 times at different locations on the sample with multiple STM tips. To get a more accurate determination of the band edges and Fermi level positions, we take a logarithm of the d$I$/d$V$ data. In doing so, we can eliminate the large suppression of d$I$/d$V$ signal close to the band edges induced by the exponential dependence of $M$ on the bias voltage $V$, i.e. $M(eV) \propto \exp(eV/\varphi)$, where $\varphi$ is an effective potential related to the work functions of the sample and the tip.[48,49] By using the fitting procedure introduced in the work by Ugeda et al.,[49] we find that SnSe$_2$ and SnSe have band gaps of $E_{SnSe_2} = 1.21 \pm 0.04$ eV and $E_{SnSe} = 1.15 \pm 0.07$ eV, respectively. These values are in close agreement with previous optical measurements.[21,22,50,51] More importantly, we find that the Fermi level in SnSe$_2$ is only 0.02 eV below the conduction band minimum (CBM), indicating a heavy n type doping reported previously by transport measurements.[23] On the other hand, the Fermi level in SnSe is about 0.1 eV above the valance band maximum (VBM), indicating a heavy p type doping which was also observed by transport.[35] These results show that SnSe$_2$-SnSe heterostructure forms a p-n junction with a type II to nearly type III band alignment. Interesting transport behavior such as rectification and negative differential conductance can be anticipated in such junctions.[52]

Besides taking point d$I$/d$V$ spectra in large flat areas of SnSe$_2$ and SnSe, we have also taken line spectra across atomically sharp interfaces. One typical result is shown in Fig. 3c, which is taken at 40 points along the dotted blue line in Fig. 2d. Clearly the gap value $E_g$ is much larger than those obtained for SnSe$_2$ and SnSe individually, and has little spatial dependence across the junction. Actually we can find that $E_g \approx E_{SnSe_2} + E_{SnSe}$. This behavior has been confirmed on other atomically sharp junctions (see SI for more data). We believe that this is due to the many nanometer sized patches of SnSe$_2$ and SnSe usually formed around the atomically sharp junctions and the special band alignment between these two materials. Charge carriers need to tunnel into either the conduction band of SnSe or the valence band of SnSe$_2$ to be able to conduct through the network of SnSe$_2$-SnSe patches to reach the STM sample plate. So the apparent band gap equals the sum of the two band gaps. While for large flat areas, they are presumably connected directly to the sample plate, allowing DOS measurements of each material independently.

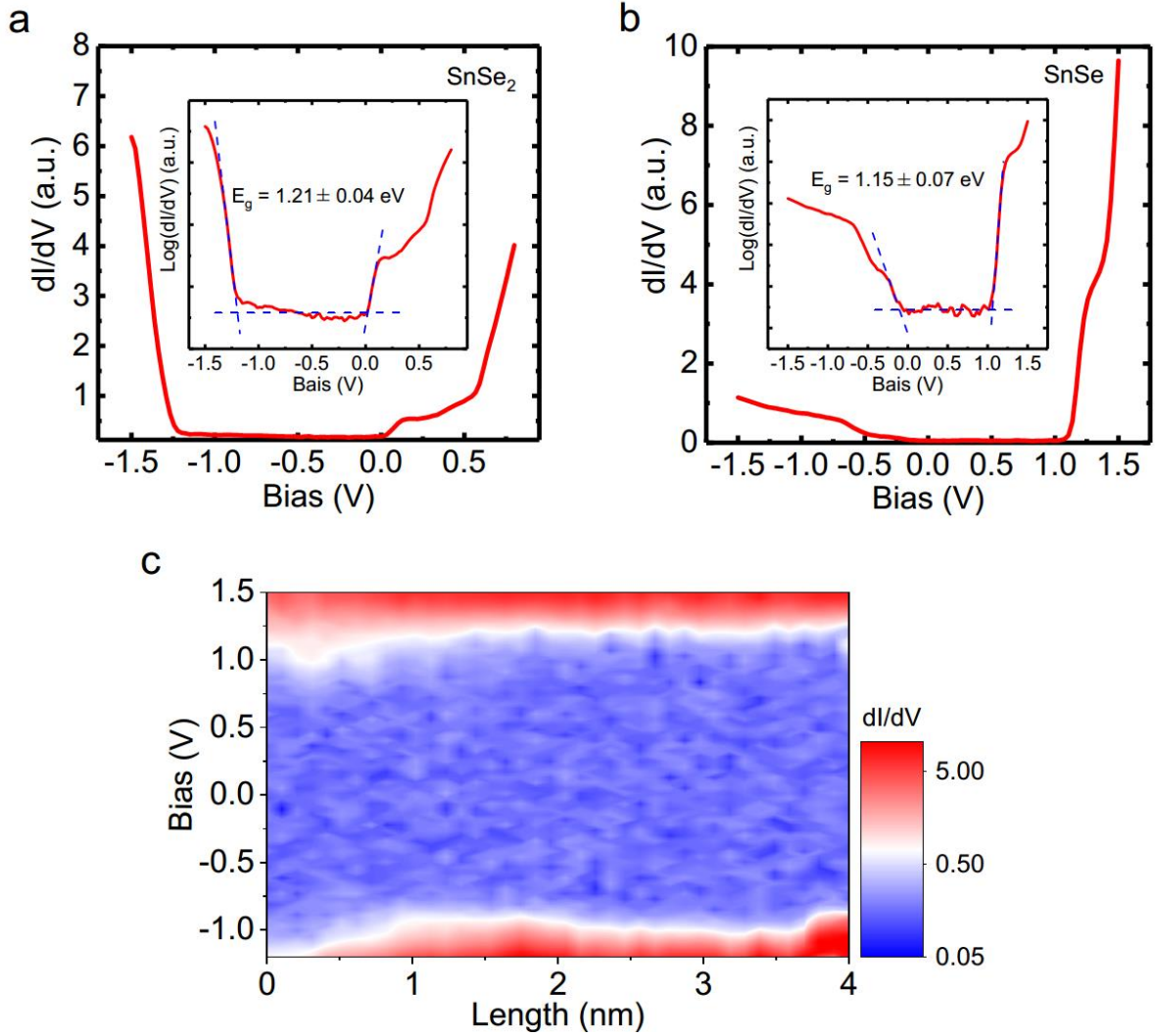

**Figure 3.** STS measurements. The d$I$/d$V$ spectra at the (a) SnSe$_2$ region, (b) SnSe region, and (c) along the blue dotted line in Fig. 2d. Insets in (a) and (b) are the corresponding semi-log plots. The dashed blue lines are the fittings to determine the band gap.[49] STS parameters: lock-in frequency $f$ = 991.2 Hz, modulation voltage $V_{rms}$ = 10 mV, and $I$ = 100 pA.

To understand the doping feature of SnSe$_2$ (SnSe), and the band alignment of the lateral SnSe$_2$-SnSe junction, we have performed systematic DFT calculations. The DOS curves of pristine SnSe$_2$ and SnSe are illustrated in Figs. 4a and d. The electronic band structures of them are provided in the SI. The calculated indirect band gaps of SnSe$_2$ and SnSe are 0.69 eV and 0.61 eV, respectively. They are smaller than the experimental results (~1.0 eV), because the Perdew–Burke-Ernzerhof (PBE) functional underestimates the band gaps of many insulators and semiconductors. But the topologies of the band structures can be well described.[53] Defects such as vacancy Se ($V_{Se}$) and interstitial Sn ($Sn_i$) in $SnSe_2$, vacancy Sn ($V_{Sn}$) and interstitial Se ($Se_i$) in SnSe have been considered by us. The DOS curves of them are shown in Fig. 4. We can clearly see that $V_{Se}$ and $Sn_i$ in $SnSe_2$ could lead to the n type doping of $SnSe_2$, while $Se_i$ and $V_{Sn}$ in SnSe would result in the p type doping of SnSe. The formation energies ($E_f$'s) of these four kinds of defects are also calculated and shown in Fig. 4g. In $SnSe_2$, the $E_f$'s of $Sn_i$ and $V_{Se}$ increase with the increasing $\mu_{Se}$, and $Sn_i$ is energetically more favored than $V_{se}$, within the whole range of chemical potential of Se ($\mu_{Se}$). In SnSe, the $E_f$'s of $V_{Sn}$ and $Se_i$ decrease with the increasing $\mu_{Se}$, and the $V_{Sn}$ is energetically more favored than $Se_i$. So the $Sn_i$ ($V_{Sn}$) should play a dominating role in determining the heavy n (p) type doping feature of $SnSe_2$ (SnSe). Besides these bulk defects, we have also calculated the $E_f$'s of the $Sn_i$ and $V_{Sn}$ located on the top surfaces of $SnSe_2$ and SnSe. The results are shown in Fig. 4g, which suggests that the $Sn_i$ on the top surface is energetically less favored than $Sn_i$ in the bulk material of $SnSe_2$. In contrast, the $V_{Sn}$ is more likely to stay on the surface of SnSe. These results nicely explain the observed topography features in Figs. 1 and 2. For SnSe in Fig. 2b, we can see missing atoms due to $V_{Sn}$ with an area density $\sim 3 \times 10^{12}$ cm$^{-2}$ (defect densities with the same order of magnitude have been observed in other scans; see Table S2). While for $SnSe_2$, despite the large electron density measured by transport, the STM topography images rarely show any defects on the surface.

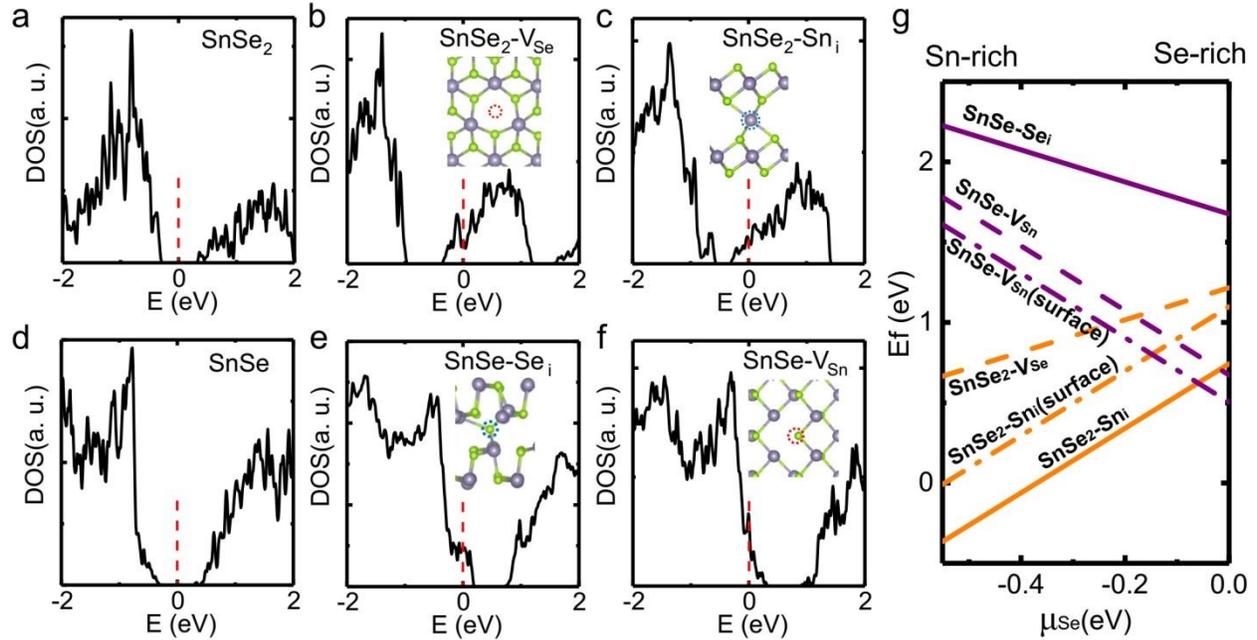

**Figure 4.** DFT calculations of DOS curves and defect formation energies. DOS curves of (a) pristine $SnSe_2$, (b) $V_{Se}$ in $SnSe_2$, (c) $Sn_i$ in $SnSe_2$, (d) pristine SnSe, (e) $Se_i$ in SnSe and (f) $V_{Sn}$ in SnSe. The vertical dashed lines indicate the positions of the Fermi levels. The atomic structures of various defects are also shown as insets. Green and violet balls are the Se and Sn atoms, respectively. Red and blue dotted circles indicate the vacancy and interstitial sites, respectively. (g) The formation energies of various defects as a function of the chemical potential of Se. Orange solid and dashed lines indicate the $E_f$'s of $Sn_i$ and $V_{Se}$ in $SnSe_2$; purple solid and dashed lines represent the $E_f$'s of $Se_i$ and $V_{Sn}$ in SnSe. The orange and purple dash-dot lines represent the $E_f$'s of $Sn_i$ and $V_{Sn}$ on the $SnSe_2$ and SnSe surfaces, respectively.

Our DFT simulation of the 2D contours of the electron partial charge density in (Fig. 5a) shows that the STM topography of SnSe only contains the tin atom positions, which is consistent with our STM topographies in Fig. 2 and a previous report.[31] In Fig. 5b the simulation for $SnSe_2$-SnSe interface well reproduces the experimental STM features in Fig. 2d. In our experiments, SnSe$_2$-SnSe heterostructures with various rotation angles have been observed. Here, we have considered three kinds of SnSe$_2$-SnSe junctions with rotation angles of 0, 15 and 30 degrees. The atomic structures of SnSe$_2$-SnSe junctions after relaxation are shown in Figs. 5c-e. For the SnSe$_2$-SnSe junction with the rotation angle of 0 degree, the interface is formed by the $(1\bar{1}0)$ surface of SnSe$_2$ and (100) surface of SnSe. For rotation angles of 15 and 30 degrees, the interface is formed respectively by the $(\bar{4}30)$ and $(2\bar{1}0)$ surfaces of SnSe$_2$ and the (100) surface of SnSe. The supercell sizes can be seen in the Figs. 5c-e. The lattice-constant mismatches between SnSe$_2$ and SnSe are 9.5%, 8.9% and 4.9% for SnSe$_2$-SnSe junctions with rotation angles of 0, 15 and 30 degrees, respectively. We note that these proposed geometries are only promising candidate structures for the SnSe$_2$-SnSe interfaces. A more systematic search of the interface atomic structures including possible interface reconstructions is out of the scope of this study. The DOS curves of these SnSe$_2$-SnSe junctions are illustrated in the right panels of Figs. 5c-e. Red and black lines indicate the DOS's of SnSe$_2$ and SnSe, respectively. All three SnSe$_2$-SnSe junctions show the type-II band alignment, with the CBM of SnSe$_2$ lying close to the VBM of SnSe, which agree well with our STS measurements.

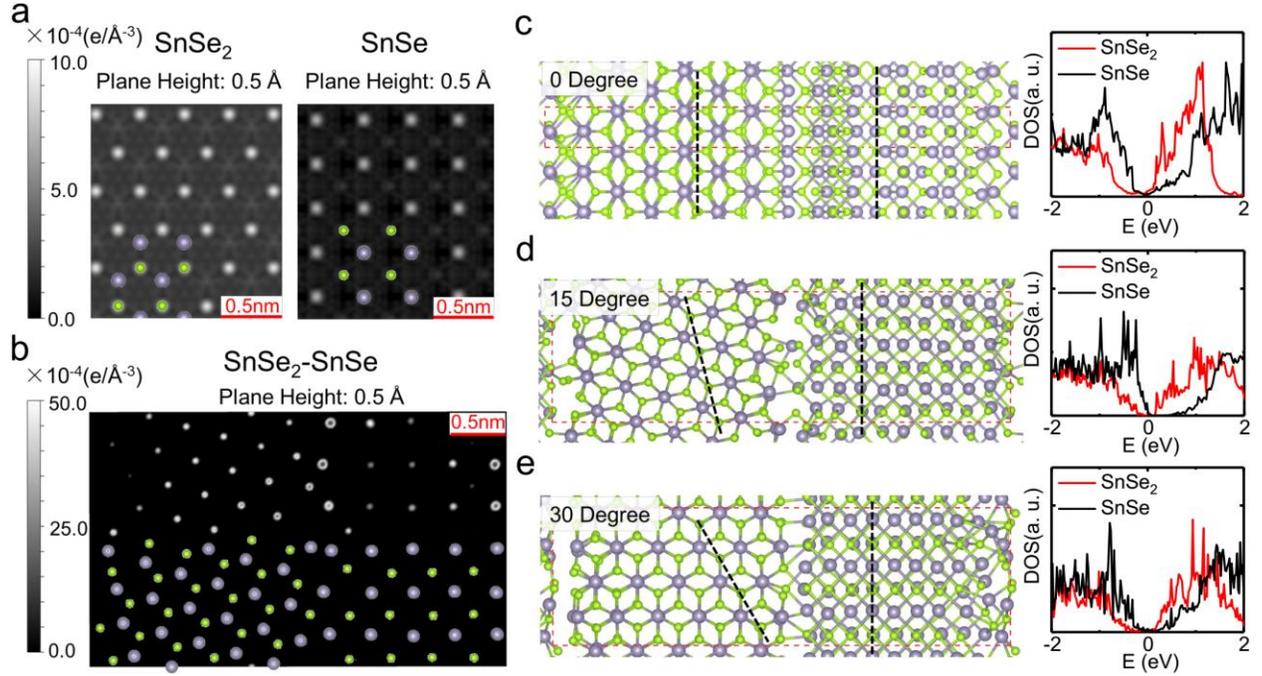

**Figure 5.** DFT calculations of the in-plane heterostructure. DFT simulation results of the 2D contours of the electron partial charge density for the slab models of (a) $SnSe_2$, SnSe and (b) $SnSe_2$-SnSe interface. The green and violet balls are Se and Sn atoms nearest to the top surface. The atomic structures and DOS curves of various $SnSe_2$-SnSe interfaces with the rotation angles between $SnSe_2$ and SnSe lattices of (c) 0, (d) 15 and (e) 30 degrees. The red dashed lines mark the supercells used in our simulations. The black dashed lines are guide for this rotation angles. For the DOS curves of $SnSe_2$ and SnSe, the Fermi levels are set to be 0 eV.

Through the comprehensive STM/STS and DFT study at the atomic scale, we have obtained the structural and electronic information of the randomly formed nanometer-sized $SnSe_2$-SnSe heterostructures. However, to further explore its properties by other macroscopic techniques, such as electronic transport measurements, we need to be able to fabricate few-layer $SnSe_2$-SnSe junctions at the micrometer scale and more importantly, at predefined locations. At first glance this may seem to be impossible by thermal annealing since the whole $SnSe_2$ sample is heated uniformly. But we find a way to overcome this difficulty. By covering part of a $SnSe_2$ flake with hexagonal boron nitride (hBN), we can protect the $SnSe_2$ underneath from transforming to SnSe, while the exposed part will be converted. Fig. 6a shows such a sample under an optical microscope. The dotted lines outline the edges of the few-layer $SnSe_2$ flake on a 300 nm $SiO_2$ substrate. The part enclosed by the green dotted line has an hBN transferred on top (see ES for fabrication details). Fig. 6b is the atomic force microscope (AFM) image of the sample before annealing, and Fig. 6c is the AFM image after annealing (see ES for annealing details). The thickness of the exposed $SnSe_2$ is reduced by 4 nm, indicating that some material has been evaporated. Then Raman spectra (Fig. 6d) are taken at the exposed and covered parts (labeled as positions 1 and 2 in Fig. 6c) of the $SnSe_2$ flake. Compared with previously reported Raman spectra for single crystalline SnSe and $SnSe_2$,[44,45] the data in Fig. 6d very nicely demonstrates that the exposed part has been fully converted to SnSe while the covered part remains as $SnSe_2$ (The hBN does not have Raman peaks in this wavenumber range. See Fig. S4 for details.). Raman mapping at the characteristic phonon modes of SnSe ($A_g^1$ mode at 65 cm$^{-1}$) and $SnSe_2$ ($A_{1g}$ mode at 182 cm$^{-1}$) unambiguously show the spatial extension of SnSe and $SnSe_2$, which is defined by the coverage of hBN. This facile method of creating $SnSe_2$-SnSe p-n junctions on demand is crucial for device fabrication and measurement.

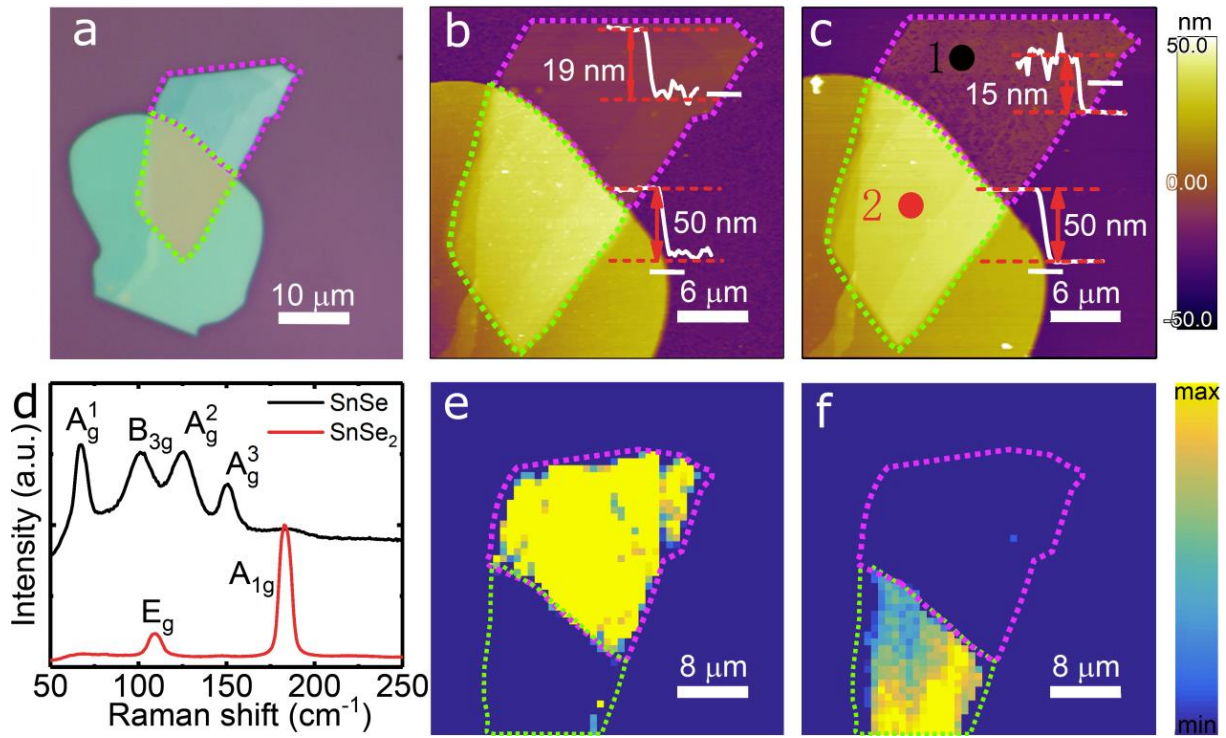

**Figure 6.** The fabrication of a SnSe$_2$-SnSe p-n junction at a predefined location. (a) The optical microscope image of a few-layer SnSe$_2$ flake partially covered by hBN. The region enclosed by the green (pink) dotted line is SnSe$_2$ with (without) hBN. The same color coding applies to (b), (c), (e) and (f). (b) and (c) AFM images of the sample shown in (a) before and after annealing. (d) Raman spectra taken at the points of 1 and 2 in (c). (e) and (f) Raman mapping of the annealed sample at the $A_g^1$ mode of SnSe and the $A_{1g}$ mode of SnSe$_2$, respectively.

CONCLUSIONS

In summary, we have proposed a novel method for creating in-plane p-n junctions in the layered material SnSe$_2$. Through comprehensive STM and STS study, we have found that thermal annealing can partially convert heavily n-type doped SnSe$_2$ to heavily p-type doped SnSe. The two lattices with distinctly different symmetries and lattice constants could form atomically sharp interfaces. DFT calculations have shown that the major dopants for SnSe$_2$ and SnSe are interstitial Sn and Sn vacancies, respectively. Calculated lattice structures and band alignments at the SnSe$_2$-SnSe junctions corroborate the experimental results. We further demonstrated the possibility of converting a predefined part of a few-layer SnSe$_2$ flake to SnSe through thermal annealing. Our findings open the door for future studies of this novel in-plane p-n junction system based on 2D materials.

EXPERIMENTAL SECTION

**Single Crystalline SnSe$_2$ Growth.** SnSe$_2$ crystals were grown by using Se as the flux. A mixture of Sn (99.999%, Aladdin) and Se granules (99.999%, Aladdin) in a molar ratio of 1:4 was placed into an alumina crucible. The crucible was sealed into a quartz tube in vacuum and was then heated in a furnace up to 800 $^{\circ}$C in 8 hours. After reaction at this temperature for 10 hours, the assembly was slowly cooled down to 400 $^{\circ}$C at a rate of 2.5 $^{\circ}$C/h and stayed at 400 $^{\circ}$C for more than 10 hours. The excess Se was quickly removed at this temperature in a centrifuge. Typical size of the crystals is 4 mm ×5 mm ×0.1 mm.

**STM Sample Preparation and Scanning Setup.** The partially converted SnSe$_2$ is obtained by annealing bulk single crystalline SnSe$_2$ (8 hours at 100℃, 12 hours at 200℃ and then 8 hours at 303℃) in the preparation chamber of an Omicron LT STM with a base pressure of $1 \times 10^{-9}$ mbar. After cooling down naturally, the sample is transferred to the LT chamber to start the STM scanning at 77.6 K.

**DFT Calculations.** The DFT calculations were performed by using the **Vienna Ab initio simulation package** (VASP) code[54,55] with the PBE functionals.[56] The projector augmented wave (PAW) potentials were used with a cutoff energy of 500 eV. A force less than 0.02 eV/Å acting on each atom was used as the criterion of relaxation. We used the 3×3×2 supercell and the 9×9×7 Monkhorst-Pack (M-P) mesh to calculate the density of states (DOS) curves in Fig. 4a-f. The formation energies ($E_f$'s) of various defects in SnSe$_2$ or SnSe were calculated by $E_f = E_{defect} - E_{SnSe_2/SnSe} - N_{Sn} \times \mu_{Sn} - N_{Se} \times \mu_{Se}$,[57,58] where $E_{defect}$ is the total energy of defect structures, $E_{SnSe_2/SnSe}$ is the total energies of pristine SnSe$_2$ or SnSe, $N_{Sn}$ and $N_{Se}$ are the number changes of Sn and Se atoms when forming the defects. The chemical potentials of Sn ($\mu_{Sn}$) and Se ($\mu_{Se}$) are tunable, but restricted by $\mu_{Sn} + 2\mu_{Se} = \Delta H_{SnSe_2}$, where $\Delta H_{SnSe_2}$ is the formation enthalpy of bulk SnSe$_2$. We employ the six-layer slab models to calculate the $E_f$'s of surface defects in Fig. 4g and plot the 2D contours of the electron partial charge density in Fig. 5a and b. The thickness of vacuum slab layer is 15 Å. The energy ranges used for the evaluation of partial charge densities are from the Fermi level (0.0 eV) to 0.5 eV.[59]

**SnSe$_2$/hBN Structure Fabrication.** At first, SnSe$_2$ flakes are exfoliated on the SiO$_2$/Si substrate. hBN flakes are exfoliated on a separate O$_2$ plasma treated SiO$_2$/Si substrate (the oxygen plasma treatment is to turn the SiO$_2$ surface to hydrophilic to facilitate later transfer steps). Then polymethyl methacrylate (PMMA) is spin-coated on it. With the assistance of water, a piece of polydimethylsiloxane (PDMS) is used to lift up the PMMA/hBN stack. Then the hBN and SnSe$_2$ are aligned under an optical microscope. Next, put the PMMA/hBN down on top of the SnSe$_2$ flake and bake at 60℃ for 2 minutes. Then take the PDMS off and bake the PMMA/hBN/SnSe$_2$ at 150℃ for 10 minutes. Later the sample is put into glacier acetic acid for 12 hours to dissolve the PMMA. Finally, the hBN/SnSe$_2$ structure is formed.

**Recipe for Annealing hBN/SnSe$_2$.** Load the sample with hBN/SnSe$_2$ into a tube furnace and pump down to $1\times10^{-5}$ torr. Raise the temperature to 300℃ at a rate of 6℃ per minute. The furnace is kept at this temperature for 200 minutes before it is cooled down naturally.

ASSOCIATED CONTENT

**Supporting Information**

The Supporting Information is available free of charge on the ACS Publications website http://pubs.acs.org.

One-atomic-layer-step interfaces between SnSe and SnSe$_2$ and their STS data; the relationship between SnSe$_2$ crystal and the thermally-converted SnSe crystal orientations; the calculated electronic band structures of bulk SnSe$_2$ and SnSe; DFT results of the lattice parameters of bulk SnSe and SnSe$_2$ crystals.

AUTHOR INFORMATION

**Corresponding Author**

*xuejm@shanghaitech.edu.cn

*yexinfeng@pku.edu.cn

**Author Contributions**

†These authors contributed equally to this work.

ACKNOWLEDGMENTS

Z.T., M.Z., C.G. and J.X. are supported by the National Natural Science Foundation of China (No. 11504234), Science and Technology Commission of Shanghai Municipality (No. 15QA1403200), the Ministry of Science and Technology of China (No. 2016YFA0204000), Shanghai Municipal Education Commission (No. 15ZZ115) and ShanghaiTech University. X.X. and Y.F. are supported by the National Natural Science Foundation of China (Nos. 11604092 and 11634001) and the National Basic Research Programs of China (No. 2016YFA0300900). W.X. and Y.G. are supported by the Natural Science Foundation of Shanghai (No. 17ZR1443300) and the Shanghai Pujiang Program (No. 17PJ1406200).

The computational resources were provided by the computation platform of National Supper-Computer Center in Changsha, China.

REFERENCES

1      Zant, P. V. *Microchip Fabrication*; McGraw-Hill Education, 2004; pp 281-287.

2      Yu, X.; Cheng, H.; Zhang, M.; Zhao, Y.; Qu, L.; Shi, G. Graphene-Based Smart Materials. *Nat. Rev. Mater.* **2017**, 2, 17046.

3      Manzeli, S.; Ovchinnikov, D.; Pasquier, D.; Yazyev, O. V.; Kis, A. 2D Transition Metal Dichalcogenides. *Nat. Rev. Mater.* **2017**, 2, 17033.

4      Novoselov, K. S.; Mishchenko, A.; Carvalho, A.; Castro Neto, A. H. 2D Materials and van der Waals Heterostructures. *Science* **2016**, 353, aac9439.

5      Xi, X.; Berger, H.; Forro, L.; Shan, J.; Mak, K. F. Gate Tuning of Electronic Phase Transitions in Two-Dimensional $NbSe_2$. *Phys. Rev. Lett.* **2016**, 117, 106801.

6   Pan, J.; Guo, C.; Song, C.; Lai, X.; Li, H.; Zhao, W.; Zhang, H.; Mu, G.; Bu, K.; Lin, T.; Xie, X.; Chen, M.; Huang, F. Enhanced Superconductivity in Restacked $TaS_2$ Nanosheets. *J. Am. Chem. Soc.* **2017**, 139, 4623-4626.

7   Ohta, T.; Bostwick, A.; Seyller, T.; Horn, K.; Rotenberg, E. Controlling The Electronic Structure of Bilayer Graphene. *Science* **2006**, 313, 951-954.

8   Cui, X.; Lee, G. H.; Kim, Y. D.; Arefe, G.; Huang, P. Y.; Lee, C. H.; Chenet, D. A.; Zhang, X.; Wang, L.; Ye, F.; Pizzocchero, F.; Jessen, B. S.; Watanabe, K.; Taniguchi, T.; Muller, D. A.; Low, T.; Kim, P.; Hone, J. Multi-Terminal Transport Measurements of $MoS_2$ Using A van der Waals Heterostructure Device Platform. *Nat. Nanotechnol.* **2015**, 10, 534-540.

9   Novoselov, K. S.; Geim, A. K.; Morozov, S. V.; Jiang, D.; Zhang, Y.; Dubonos, S. V.; Grigorieva, I. V.; Firsov, A. A. Electric Field Effect in Atomically Thin Carbon Films. *Science* **2004**, 306, 666-669.

10  Chu, T.; Ilatikhameneh, H.; Klimeck, G.; Rahman, R.; Chen, Z. Electrically Tunable Bandgaps in Bilayer $MoS_2$. *Nano. Lett.* **2015**, 15, 8000-8007.

11  Zeng, H.; Dai, J.; Yao, W.; Xiao, D.; Cui, X. Valley Polarization in $MoS_2$ Monolayers by Optical Pumping. *Nat. Nanotechnol.* **2012**, 7, 490-493.

12  Withers, F.; Del Pozo-Zamudio, O.; Schwarz, S.; Dufferwiel, S.; Walker, P. M.; Godde, T.; Rooney, A. P.; Gholinia, A.; Woods, C. R.; Blake, P.; Haigh, S. J.; Watanabe, K.; Taniguchi, T.; Aleiner, I. L.; Geim, A. K.; Fal'ko, V. I.; Tartakovskii, A. I.; Novoselov, K. S. $WSe_2$ Light-Emitting Tunneling Transistors with Enhanced Brightness at Room Temperature. *Nano. Lett.* **2015**, 15, 8223-8228.

13  Burg, G. W.; Prasad, N.; Fallahazad, B.; Valsaraj, A.; Kim, K.; Taniguchi, T.; Watanabe, K.; Wang, Q.; Kim, M. J.; Register, L. F.; Tutuc, E. Coherent Interlayer Tunneling and Negative Differential Resistance with High Current Density in Double Bilayer Graphene-WSe$_2$ Heterostructures. *Nano. Lett.* **2017**, 17, 3919-3925.

14  Long, M.; Liu, E.; Wang, P.; Gao, A.; Xia, H.; Luo, W.; Wang, B.; Zeng, J.; Fu, Y.; Xu, K.; Zhou, W.; Lv, Y.; Yao, S.; Lu, M.; Chen, Y.; Ni, Z.; You, Y.; Zhang, X.; Qin, S.; Shi, Y.; Hu, W.; Xing, D.; Miao, F. Broadband Photovoltaic Detectors Based on An Atomically Thin Heterostructure. *Nano. Lett.* **2016**, 16, 2254-2259.

15  Duan, X.; Wang, C.; Shaw, J. C.; Cheng, R.; Chen, Y.; Li, H.; Wu, X.; Tang, Y.; Zhang, Q.; Pan, A.; Jiang, J.; Yu, R.; Huang, Y.; Duan, X. Lateral Epitaxial Growth of Two-Dimensional Layered Semiconductor Heterojunctions. *Nat. Nanotechnol.* **2014**, 9, 1024-1030.

16  Gong, Y.; Lin, J.; Wang, X.; Shi, G.; Lei, S.; Lin, Z.; Zou, X.; Ye, G.; Vajtai, R.; Yakobson, B. I.; Terrones, H.; Terrones, M.; Tay, B. K.; Lou, J.; Pantelides, S. T.; Liu, Z.; Zhou, W.; Ajayan, P. M. Vertical and In-Plane Heterostructures from WS$_2$/MoS$_2$ Monolayers. *Nat. Mater.* **2014**, 13, 1135-1142.

17  Huang, C.; Wu, S.; Sanchez, A. M.; Peters, J. J.; Beanland, R.; Ross, J. S.; Rivera, P.; Yao, W.; Cobden, D. H.; Xu, X. Lateral Heterojunctions within Monolayer MoSe$_2$-WSe$_2$ Semiconductors. *Nat. Mater.* **2014**, 13, 1096-1101.

18  Li, M. Y.; Shi, Y.; Cheng, C. C.; Lu, L. S.; Lin, Y. C.; Tang, H. L.; Tsai, M. L.; Chu, C. W.; Wei, K. H.; He, J. H.; Chang, W. H.; Suenaga, K.; Li, L. J. Epitaxial Growth of A Monolayer WSe$_2$-MoS$_2$ Lateral p-n Junction with An Atomically Sharp Interface. *Science* **2015**, 349, 524-528.

19   Mahjouri-Samani, M.; Lin, M. W.; Wang, K.; Lupini, A. R.; Lee, J.; Basile, L.; Boulesbaa, A.; Rouleau, C. M.; Puretzky, A. A.; Ivanov, I. N.; Xiao, K.; Yoon, M.; Geohegan, D. B. Patterned Arrays of Lateral Heterojunctions within Monolayer Two-Dimensional Semiconductors. *Nat. Commun.* **2015**, 6, 7749.

20   Zhang, Z.; Chen, P.; Duan, X.; Zang, K.; Luo, J.; Duan, X. Robust Epitaxial Growth of Two-Dimensional Heterostructures, Multiheterostructures, and Superlattices. *Science* **2017**, 357, 788-792.

21   Domingo, G.; Itoga, R. S.; Kannewurf, C. R. Fundamental Optical Absorption in $SnS_2$ and $SnSe_2$. *Phys. Rev.* **1966**, 143, 536-541.

22   Lee, P. A.; Said, G. Optical Properties of Tin Di-Selenide Single Crystals. *J. Phys. D: Appl. Phys.* **1968**, 1, 837-843.

23   Guo, C.; Tian, Z.; Xiao, Y.; Mi, Q.; Xue, J. Field-Effect Transistors of High-Mobility Few-Layer $SnSe_2$. *Appl. Phys. Lett.* **2016**, 109, 203104.

24   Zhou, X.; Gan, L.; Tian, W.; Zhang, Q.; Jin, S.; Li, H.; Bando, Y.; Golberg, D.; Zhai, T. Ultrathin $SnSe_2$ Flakes Grown by Chemical Vapor Deposition for High-Performance Photodetectors. *Adv. Mater.* **2015**, 27, 8035-8041.

25   Asanabe, S. Electrical Properties of Stannous Selenide. *J. Phys. Soc. Japan* **1959**, 14, 281-296.

26   Peters, M. J.; McNeil, L. E. High-Pressure Mössbauer Study of SnSe. *Phys. Rev. B* **1990**, 41, 5893-5897.

27   Zhao, L. D.; Lo, S. H.; Zhang, Y.; Sun, H.; Tan, G.; Uher, C.; Wolverton, C.; Dravid, V. P.; Kanatzidis, M. G. Ultralow Thermal Conductivity and High Thermoelectric Figure of Merit in SnSe Crystals. *Nature* **2014**, 508, 373-377.

28  Zhao, L. D.; Tan, G.; Hao, S.; He, J.; Pei, Y.; Chi, H.; Wang, H.; Gong, S.; Xu, H.; Dravid, V. P.; Uher, C.; Snyder, G. J.; Wolverton, C.; Kanatzidis, M. G. Ultrahigh Power Factor and Thermoelectric Performance in Hole-Doped Single-Crystal SnSe. *Science* **2016**, 351, 141-144.

29  Li, C. W.; Hong, J.; May, A. F.; Bansal, D.; Chi, S.; Hong, T.; Ehlers, G.; Delaire, O. Orbitally Driven Giant Phonon Anharmonicity in SnSe. *Nat. Phys.* **2015**, 11, 1063-1069.

30  Kim, S.-u.; Duong, A.-T.; Cho, S.; Rhim, S. H.; Kim, J. A Microscopic Study Investigating The Structure of SnSe Surfaces. *Surf. Sci.* **2016**, 651, 5-9.

31  Duvjir, G.; Min, T.; Thi Ly, T.; Kim, T.; Duong, A.-T.; Cho, S.; Rhim, S. H.; Lee, J.; Kim, J. Origin of p-Type Characteristics in A SnSe Single Crystal. *Appl. Phys. Lett.* **2017**, 110, 262106.

32  Lu, Q.; Wu, M.; Wu, D.; Chang, C.; Guo, Y. P.; Zhou, C. S.; Li, W.; Ma, X. M.; Wang, G.; Zhao, L. D.; Huang, L.; Liu, C.; He, J. Unexpected Large Hole Effective Masses in SnSe Revealed by Angle-Resolved Photoemission Spectroscopy. *Phys. Rev. Lett.* **2017**, 119, 116401.

33  Wang, C. W.; Xia, Y. Y. Y.; Tian, Z.; Jiang, J.; Li, B. H.; Cui, S. T.; Yang, H. F.; Liang, A. J.; Zhan, X. Y.; Hong, G. H.; Liu, S.; Chen, C.; Wang, M. X.; Yang, L. X.; Liu, Z.; Mi, Q. X.; Li, G.; Xue, J. M.; Liu, Z. K.; Chen, Y. L. Photoemission Study of The Electronic Structure of Valence Band Convergent SnSe. *Phys. Rev. B* **2017**, 96, 165118.

34  Zhao, S.; Wang, H.; Zhou, Y.; Liao, L.; Jiang, Y.; Yang, X.; Chen, G.; Lin, M.; Wang, Y.; Peng, H.; Liu, Z. Controlled Synthesis of Single-Crystal SnSe Nanoplates. *Nano Res.* **2015**, 8, 288-295.

35  Pei, T.; Bao, L.; Ma, R.; Song, S.; Ge, B.; Wu, L.; Zhou, Z.; Wang, G.; Yang, H.; Li, J.; Gu, C.; Shen, C.; Du, S.; Gao, H.-J. Epitaxy of Ultrathin SnSe Single Crystals on Polydimethylsiloxane: In-Plane Electrical Anisotropy and Gate-Tunable Thermopower. *Adv. Electron. Mater.* **2016**, 2, 1600292.

36  Shi, G.; Kioupakis, E. Anisotropic Spin Transport and Strong Visible-Light Absorbance in Few-Layer SnSe and GeSe. *Nano. Lett.* **2015**, 15, 6926-6931.

37  Fei, R.; Li, W.; Li, J.; Yang, L. Giant Piezoelectricity of Monolayer Group IV Monochalcogenides: SnSe, SnS, GeSe, and GeS. *Appl. Phys. Lett.* **2015**, 107, 173104.

38  Gomes, L. C.; Carvalho, A. Phosphorene Analogues: Isoelectronic Two-Dimensional Group-IV Monochalcogenides with Orthorhombic Structure. *Phys. Rev. B* **2015**, 92, 085406.

39  Zhang, L. C.; Qin, G.; Fang, W. Z.; Cui, H. J.; Zheng, Q. R.; Yan, Q. B.; Su, G. Tinselenidene: A Two-Dimensional Auxetic Material with Ultralow Lattice Thermal Conductivity and Ultrahigh Hole Mobility. *Sci. Rep.* **2016**, 6, 19830.

40  Rodin, A. S.; Gomes, L. C.; Carvalho, A.; Castro Neto, A. H. Valley Physics in Tin (II) Sulfide. *Phys. Rev. B* **2016**, 93, 045431.

41  Wu, M.; Zeng, X. C. Intrinsic Ferroelasticity and/or Multiferroicity in Two-Dimensional Phosphorene and Phosphorene Analogues. *Nano. Lett.* **2016**, 16, 3236-3241.

42  Sutter, E.; Huang, Y.; Komsa, H. P.; Ghorbani-Asl, M.; Krasheninnikov, A. V.; Sutter, P. Electron-Beam Induced Transformations of Layered Tin Dichalcogenides. *Nano. Lett.* **2016**, 16, 4410-4416.

43  Pałosz, B.; Salje, E. Lattice Parameters and Spontaneous Strain in $AX_2$ Polytypes: $CdI_2$, $PbI_2$, $SnS_2$ and $SnSe_2$. *J. Appl. Cryst.* **1989**, 22, 622-623.

44    Smith, A. J.; Meek, P. E.; Liang, W. Y. Raman Scattering Studies of $SnS_2$ and $SnSe_2$. *J. Phys. C: Solid State Phys.* **1977**, 10, 1321-1333.

45    Chandrasekhar, H. R.; Humphreys, R. G.; Zwick, U.; Cardona, M. Infrared and Raman Spectra of The IV-VI Compounds SnS and SnSe. *Phys. Rev. B* **1977**, 15, 2177-2183.

46    Maier, H.; Daniel, D. R. SnSe Single Crystals: Sublimation Growth, Deviation from Stoichiometry and Electrical Properties. *J. Electron. Mater.* **1977**, 6, 693-704.

47    Chattopadhyay, T.; Pannetier, J.; Vonschnering, H. G. Neutron-Diffraction Study of The Structural Phase-Transition in SnS and SnSe. *J. Phys.Chem. Solids* **1986**, 47, 879–885.

48    Chen, C. J. *Introduction to Scanning Tunneling Microscopy*; Oxford Univ. Press, 1993; pp 68-69.

49    Ugeda, M. M.; Bradley, A. J.; Shi, S. F.; da Jornada, F. H.; Zhang, Y.; Qiu, D. Y.; Ruan, W.; Mo, S. K.; Hussain, Z.; Shen, Z. X.; Wang, F.; Louie, S. G.; Crommie, M. F. Giant Bandgap Renormalization and Excitonic Effects in A Monolayer Transition Metal Dichalcogenide Semiconductor. *Nat. Mater.* **2014**, 13, 1091-1095.

50    Albers, W.; Haas, C.; Ober, H.; Sehrodder, G. R.; Wasscher, J. D. Preparation and Properties of Mixed Crystals $SnS_{(1-x)}Se_x$. *J. Phys. Chem. Solids* **1962**, 23, 215-220.

51    Madelung, O. *Semiconductors: Data Handbook*; Springer Press, 2014; pp 579-603.

52    Yan, R.; Fathipour, S.; Han, Y.; Song, B.; Xiao, S.; Li, M.; Ma, N.; Protasenko, V.; Muller, D. A.; Jena, D.; Xing, H. G. Esaki Diodes in van der Waals Heterojunctions with Broken-Gap Energy Band Alignment. *Nano. Lett.* **2015**, 15, 5791-5798.

53    Gonzalez, J. M.; Oleynik, I. I. Layer-Dependent Properties of $SnS_2$ and $SnSe_2$ Two-Dimensional Materials. *Phys. Rev. B* **2016**, 94, 125443.

54      Kresse, G.; Hafner, J. Ab Initiomolecular Dynamics for Liquid Metals. *Phys. Rev. B* **1993**, 47, 558-561.

55      Kresse, G.; Hafner, J. Ab Initiomolecular-Dynamics Simulation of The Liquid-Metal–Amorphous-Semiconductor Transition in Germanium. *Phys. Rev. B* **1994**, 49, 14251-14269.

56      Perdew, J. P.; Burke, K.; Ernzerhof, M. Generalized Gradient Approximation Made Simple. *Phys. Rev. Lett.* **1996**, 77, 3865-3868.

57      Zhang, S. B.; Northrup, J. E. Chemical Potential Dependence of Defect Formation Energies in GaAs: Application to Ga Self-Diffusion. *Phys. Rev. Lett.* **1991**, 67, 2339-2342.

58      Kumagai, Y.; Burton, L. A.; Walsh, A.; Oba, F. Electronic Structure and Defect Physics of Tin Sulfides: $SnS, Sn_2S_3$, and $SnS_2$. *Phys. Rev. Applied.* **2016**, 6, 014009.

59      Lin, C.; Feng, Y.; Xiao, Y.; Durr, M.; Huang, X.; Xu, X.; Zhao, R.; Wang, E.; Li, X. Z.; Hu, Z. Direct Observation of Ordered Configurations of Hydrogen Adatoms on Graphene. *Nano. Lett.* **2015**, 15, 903-908.